\documentstyle[11pt,epsf,rotating]{article}
\newcommand{\zp}[3]{Z. Phys.\ C#1 (19#2) #3}
\newcommand{\pl}[3]{Phys.\ Lett.\ B#1 (19#2) #3}
\newcommand{\np}[3]{Nucl.\ Phys.\ B#1 (19#2) #3}
\newcommand{\prd}[3]{Phys.\ Rev.\ D#1 (19#2) #3}

\newcommand{\niam}[3]{Nucl.\ Instr.\ and Meth.\ #1 (19#2) #3}
\newcommand{\gaga}{\mbox{$\gamma\gamma$}}

\def\simgt{\rlap{\lower 3.5 pt \hbox{$\mathchar \sim$}} \raise 1pt \hbox {$>$}}
\def\simlt{\rlap{\lower 3.5 pt \hbox{$\mathchar \sim$}} \raise 1pt \hbox {$<$}}

\newcommand{\qqb}[1]{\mbox{$#1\overline{#1}$}}

\newcommand{\beq}{\begin{equation}}
\newcommand{\eeq}{\end{equation}}
\newcommand{\bea}{\begin{eqnarray}}
\newcommand{\eea}{\end{eqnarray}}

\catcode`\@=11

\def\section{\@startsection{section}{1}{\z@}{3.5ex plus 1ex minus .2ex}
{2.3ex plus .2ex}{\large\bf}}

\def\thesection{\arabic{section}.}

\def\appendix{\setcounter{section}{0}
 \def\thesection{Appendix \Alph{section}:}
 \def\theequation{\Alph{section}.\arabic{equation}}}
\def\@citex[#1]#2{\if@filesw\immediate\write\@auxout{\string\citation{#2}}\fi
  \def\@citea{}\@cite{\@for\@citeb:=#2\do
    {\@citea\def\@citea{,\penalty\@m}\@ifundefined
       {b@\@citeb}{{\bf ?}\@warning
       {Citation `\@citeb' on page \thepage \space undefined}}%
\hbox{\csname b@\@citeb\endcsname}}}{#1}}

\def\citer{\@ifnextchar [{\@tempswatrue\@citexr}{\@tempswafalse\@citexr[]}}

\def\@citexr[#1]#2{\if@filesw\immediate\write\@auxout{\string\citation{#2}}\fi
  \def\@citea{}\@cite{\@for\@citeb:=#2\do
    {\@citea\def\@citea{--\penalty\@m}\@ifundefined
       {b@\@citeb}{{\bf ?}\@warning
       {Citation `\@citeb' on page \thepage \space undefined}}%
\hbox{\csname b@\@citeb\endcsname}}}{#1}}
\relax
\begin{document}

\thispagestyle{empty}

\hfill\vbox{\hbox{\bf DESY 95--205}
            \hbox{\bf FNT/T--95/28}
            \hbox{\bf LNF--95/059(P)}
            \hbox{\bf MPI/PhT/95--113}
            \hbox{December 1995}           }

\vspace{0.5in}
\begin{center}
\boldmath
{\Large\bf Large-$p_\perp$ Heavy-Quark Production in Two-Photon Collisions}
\unboldmath
\vspace{0.5in}

{\large \sc M.~Cacciari$^{a\dagger}$, M.~Greco$^{b}$, B.A.~Kniehl$^{c}$,
            M.~Kr\"amer$^{a}$, G.~Kramer$^{d}$, M.~Spira$^{d}$}

\vspace{0.5in}

$^a$Deutsches Elektronen-Synchrotron DESY, 22603 Hamburg, Germany\\[2mm]
$^b$Dipartimento di Fisica, Universit\`{a} di Roma III, and\\[1mm]
$\hphantom{^b}$INFN, Laboratori Nazionali di Frascati, Italy\\[2mm]
$^c$Max-Planck-Institut f\"ur Physik, F\"ohringer Ring 6, 80805 Munich,
    Germany\\[2mm]
$^d$II. Institut f\"ur Theoretische Physik$^{\ddagger}$, Universit\"at
    Hamburg, 22761 Hamburg, Germany

\end{center}

\vspace{0.5in}

\begin{abstract}
  The next-to-leading-order (NLO) cross section for the production of
  heavy quarks at large transverse momenta ($p_\perp$) in
  $\gamma\gamma$ collisions is calculated with perturbative
  fragmentation functions (PFF's).  This approach allows for a
  resummation of terms $\propto\alpha_s\ln(p_\perp^2/m^2)$ which arise
  in NLO from collinear emission of gluons by heavy quarks at large
  $p_\perp$ or from almost collinear branching of photons or gluons
  into heavy-quark pairs. We present single-inclusive distributions in
  $p_\perp$ and rapidity including direct and resolved photons for
  $\gamma\gamma$ production of heavy quarks at $e^+e^-$ colliders and
  at high-energy $\gamma\gamma$ colliders. The results are compared
  with the fixed-order calculation for $m$ finite including QCD
  radiative corrections. The two approaches differ in the definitions
  and relative contributions of the direct and resolved terms, but
  essentially agree in their sum. The resummation of the $\alpha_s
  \ln(p_\perp^2/m^2)$ terms in the PFF approach leads to a softer
  $p_\perp$ distribution and to a reduced sensitivity to the choice of
  the renormalization and factorization scales.
\end{abstract}

\vfill

\noindent
{\footnotesize
$\dagger\,$ Della Riccia and Universit\`a di Pavia fellow.
            Work also supported by INFN, Sezione di Pavia.}\\
{\footnotesize
$\ddagger\,$ Supported by Bundesministerium f\"ur Forschung und
            Technologie,
            Bonn, Germany, Contract 05 6 HH 93P (5) and EEC Program
            {\it Human Capital and Mobility} through Network {\it Physics
            at High Energy Colliders} under Contract CHRX--CT93--0357
            (DG12 COMA).}

\newpage

\section{Introduction}
A large number of equivalent photons is generated at high-energy
$e^+e^-$ colliders, giving rise to the production of heavy-quark pairs
in two-photon collisions. This process has been studied for charmed
particles by several experiments at PETRA, PEP, TRISTAN, and LEP
\cite{ggdata}. At LEP2, a total of $\sim$ 350\,000 $c\bar{c}$ and
$\sim$ 1500 $b\bar{b}$ pairs will be produced in $\gamma\gamma$
collisions for an integrated luminosity of $\int{\cal{L}} =
500$~pb$^{-1}$ \cite{dkzz}.  The yield of heavy quarks at high-energy
$e^+e^-$ linear colliders is even higher, depending in detail on the
spectrum of the beamstrahlung photons, which strongly varies with the
machine design and operation. If the novel method of Compton
back-scattering of laser light can be made work \cite{gkst}, it will
be possible to generate high-luminosity beams of real photons carrying
$\sim$~80\% of the electron/positron energy.  The high-statistics data
from the future experimental facilities will allow for a detailed
comparison of the next-to-leading order (NLO) predictions with
experimental results not only for total production rates, but also for
various differential distributions. These analyses will provide us
with information on the dynamics of heavy-flavour production in a
kinematical range very different from that available in $\gamma\gamma$
collisions at present colliders.

Three mechanisms contribute to the production of heavy quarks in
\gaga\ collisions: $(i)$ In the case of direct (DD) production, the
two photons couple directly to the heavy quarks. No spectator
particles travel along the $\gamma$~axes. $(ii)$ If one of the photons
first splits into a flux of light quarks and gluons \cite{phsplit},
one of the gluons may fuse with the second photon to form the \qqb{Q}
pair. The remaining light quarks and gluons build up a spectator jet
in the split-$\gamma$ direction (single-resolved (DR) $\gamma$
contribution). The total \gaga\ cross section of this mechanism
depends on the parton density functions (PDF's) of the photon
\citer{photstruc,acfgp}.  $(iii)$ If both photons split into light
quarks and gluons, the \qqb{Q} pair is accompanied by two spectator
jets (double-resolved (RR) $\gamma$ contribution).  Since the photon
PDF's scale as $\alpha\alpha_s^{-1}$, the DR and RR processes are of
the same order as the DD process.

Many features of the above-mentioned production mechanisms are
calculable in perturbative QCD. The mass of the heavy quark, $m \gg
\Lambda_{\mbox{\scriptsize QCD}}$, acts as a cutoff and sets the scale
for the perturbative calculations.  The production cross section
factorizes into a partonic hard-scattering cross section multiplied by
light-quark and gluon PDF's \cite{coll}. Inherent in this
factorization scheme is the notion that the only quarks in the photon
are the light ones. There are no contributing subprocesses initiated
by an intrinsic heavy flavour coming directly from the photon PDF's.
In leading order (LO), direct production is described by the partonic
reaction $\gamma + \gamma \to c + \bar{c}$ while the resolved
contributions involve the channels $g+\gamma \to c + \bar{c}$ (DR) and
$q+\bar{q}\to c+\bar{c}$, $g+g\to c+\bar{c}$ (RR), where $q$ are light
(massless) flavours \cite{dgod}.  The NLO corrections to these
processes have been calculated and found to be substantial
\cite{dkzz,nde,nde2,nlohq}.  A comparison between the NLO results and
experimental data on the total cross section of charm-quark production
in two-photon collisions has shown satisfactory agreement
\cite{mkphot95}.

One might expect that the massive approach is reasonable only in those
kinematical regions where the mass $m$ and any other characteristic
energy scale like $p_\perp$ are approximately of the same magnitude
and significantly larger than $\Lambda_{\mbox{\scriptsize QCD}}$. In
NLO, terms $\propto\alpha_s\ln(p_\perp^2/m^2)$ arise from collinear
emission of gluons by heavy quarks at large transverse momenta
($p_\perp$) or from almost collinear branching of gluons or photons
into heavy-quark pairs. These terms are not expected to affect the
total production rates, but they might spoil the convergence of the
perturbation series and cause large scale dependences of the NLO
result at $p_\perp \gg m$.\footnote{Similar potentially large terms
  $\propto\alpha_s\ln(Q^2/m^2)$, $Q$ being the photon virtuality, do
  appear in NLO calculations of heavy quark electroproduction cross
  sections. The resummation of these logarithms is being considered in
  a series of papers \cite{dis}.} In the massive approach, the
prediction of differential cross sections is thus limited to a rather
small range of $p_\perp\sim m$.  An alternative way of making
predictions at large $p_\perp$ is to treat the heavy quarks as
massless partons. The mass singularities of the form
$\ln(p_\perp^2/m^2)$ are then absorbed into the PDF's and
fragmentation functions (FF's) in the same way as for the light $u$,
$d$, and $s$ quarks. The crucial difference to the production of light
hadrons is the fact that the initial-state conditions for the
heavy-quark FF's can be calculated within perturbative QCD and do not
have to be taken from experiment \cite{MN91}.  Such perturbative
fragmentation functions (PFF's) have been been used in
Ref.~\cite{CG94} to study the production of large-$p_\perp$ bottom
quarks in $p\bar{p}$ collisions. Meanwhile, similar analyses have been
carried out for charm-quark production in photon-proton collisions at
HERA \cite{kkks,CG95}. In the massless scheme, the heavy quark is, of
course, considered to be one of the massless active flavours in the
photon PDF's. We expect the massless PFF approach to be better suited
for the calculation of the differential $p_\perp$ distributions at NLO
in the region $p_\perp \gg m$. The small-$p_\perp$ region is, however,
not calculable without retaining the full $m$ dependence. The massless
cross section diverges in the limit $p_\perp \to 0$, and total
production rates can not be predicted.

At this point, it is worth mentioning that the massive and the
massless approaches are expected to give different descriptions of the
DD, DR, and RR contributions. In the massive approach, the direct and
resolved contributions are separately well defined through NLO. In
fact, there is no need to perform any factorization for the incoming
photons. This means that an experiment analyzing data without hadronic
activity in the directions of the incoming photons would directly
probe the NLO calculation of DD heavy-quark production. This is,
however, not true in the massless approach.  If the charm is treated
as a massless parton, the photon splitting to $c\bar c$ must also
undergo a subtraction procedure.  The DD piece then becomes
factorization-scheme dependent and looses its direct physical
interpretation.  Consequently, the same is true for the DR and RR
parts.  As a matter of principle, the three contributions cannot be
experimentally separated any more with NLO accuracy.  Only their sum
corresponds to a physical observable.  On the other hand, if one were
to extend the massive approach up to large $p_\perp$, the DD
prediction would, of course, still be unambiguously defined, but it
would be affected by large logarithmic terms, which would render it
unreliable. Only at moderate $p_\perp \sim m$ and through NLO, it is
therefore possible to probe the theoretical predictions for direct and
resolved contributions separately by an experimental analysis.
%perform an experimental analysis of the different
%production mechanisms with an accuracy comparable to the one of the
%theoretical calculation.

In the near future, we expect experimental data in the intermediate
$p_\perp$ range, where $p_\perp>m$ rather than $p_\perp\gg m$.  Then,
the problem how to proceed in this $p_\perp$ region arises.  In order
to investigate the region where $p_\perp > m$, we calculate the
differential cross section
$\mbox{d}^2\sigma/\mbox{d}y\,\mbox{d}p_\perp^2$ as a function of
$p_\perp$ with fixed rapidity $y$.  We compare the results in the two
approaches: $(i)$ the massive-charm approach with $m = 1.5$~GeV, in
which we compute the cross section for open charm production, and
$(ii)$ the massless approach, where we evaluate the same differential
cross section for inclusive charm production using PFF's. In both
calculations, we include the DD, DR, and RR processes up to NLO.  The
massive calculation is based on the work presented in \cite{dkzz}, in
which the NLO theory for DD and DR production was elaborated.  The
calculation of the massive RR cross section relies on the work of
\cite{nde,nde2}.  The massless calculation proceeds along the lines of
\citer{CG94,kk} on the basis of the DR and RR hard-scattering cross
sections obtained in \cite{aurenche,aversa}.  The NLO corrections to
the DD cross section were derived in \cite{fon} and recently confirmed
in \cite{gordon}.  In the massless approach, we adopt the PFF's
calculated in \cite{MN91}.  With this choice of PFF's, the LO results
in the massive scheme approach the LO massless results in the limit $m
\to 0$, if we restrict ourselves to the same parton subprocesses.  In
LO, the PFF's are equal to $\delta(1-x)$ (see Eq.~(\ref{eqfrag}))
showing that, in LO, we have full correspondence between the massless
and massive approaches.  However, they must differ in NLO, where the
limit $m \to 0$ is not possible due to the unabsorbed mass-singular
$\ln(p_\perp^2/m^2)$ terms in the massive approach.  In the massless
approach, these terms are contained in the higher-order terms of the
photon PDF's and the PFF's.  This situation is very similar to our
previous study for photon-proton collisions at HERA \cite{kkks,CG95}.
Indeed the direct and resolved contributions in $\gamma p$ scattering
correspond to the DR and RR contributions in $\gamma\gamma$ collisions
with the proton PDF's replaced by the photon PDF's.  In addition, we
now have the DD contribution.
%In the massless approach, all these
%contributions are interrelated in NLO through factorization.

The outline of our work is as follows.  In Section~2, we shortly
describe the basic formalism for the massless PFF approach.  Section~3
contains the numerical results.  Our conclusions are summarized in
Section~4.

\section{The perturbative fragmentation function approach}

In this section, we describe in some detail the PFF approach to
heavy-quark production at large $p_\perp$. This technique was proposed
in \cite{nde2} and first applied in the context of hadron collisions
for describing large-$p_\perp$ bottom production \cite{CG94}.  The
basic assumption is that when a large scale is governing the
production process (in our case, this is $p_\perp$, with $p_\perp\gg
m$) the heavy quark itself is produced as if it was massless.
Technically speaking, non-singular mass terms are suppressed in the
cross section by powers of $m/p_\perp$. The important mass terms
appear whenever the virtuality of the heavy quark is small.  This
happens in the initial state when the heavy parton is emitted from the
colliding hadron, and in the final state when the partons materialize
into a massive quark.

This qualitative picture of heavy-quark production in the
large-$p_\perp$ limit can be substantiated at NLO in the following
way:
$(i)$ The hard-scattering cross sections are calculated in the
massless approximation, and the collinear singularities are subtracted
according to some factorization scheme, {\it e.g.}, the
$\overline{\mbox{MS}}$ scheme.  Since the heavy parton is taken to be
massless, also the singularities arising from its splittings are
subtracted.
$(ii)$ As for the initial state, the heavy parton is accommodated in
the PDF's like a light flavour.  The massiveness of the quark is
usually taken into account by including it in the evolution only above
a scale set by its mass.
$(iii)$ As for the final state, the PFF's characterize the
hadronization of the massless partons into the heavy-quark state.
Exploiting the fact that the produced quark has mass
$m\gg\Lambda_{\mbox{\scriptsize QCD}}$, universal starting conditions
for these FF's can be calculated within perturbative QCD (therefore
they are denoted PFF's) at a scale $\mu_0$ of order $m$.  In
\cite{MN91}, these starting conditions were calculated at NLO in the
$\overline{\mbox{MS}}$ scheme. They read
\begin{eqnarray}\label{eqfrag}
D_Q^Q(x,\mu_0)&=& \delta(1-x) + {{\alpha_s(\mu_0)}\over{2\pi}} C_F
\left\{{{1+x^2}\over{1-x}}\left[\log{{\mu_0^2}\over{m^2}} -2\log(1-x)
-1\right]\right\}_+ \label{DQQ}, \nonumber \\
D_g^Q(x,\mu_0)&=& {{\alpha_s(\mu_0)}\over{2\pi}} T_f
[x^2 + (1-x)^2]\log{{\mu_0^2}\over{m^2}} \label{DgQ}, \nonumber \\
D_{q,\bar q,\bar Q}^Q(x,\mu_0)&=& 0 \label{DqQ},
\end{eqnarray}
where $D_a^Q$ refers to the fragmentation of parton $a$ into the heavy
quark $Q$, $C_F=4/3$, and $T_f=1/2$.
$(iv)$~Finally, the PDF's and PFF's are evolved in NLO up to the
chosen factorization scale (which is usually of the order of
$p_\perp$) via the Altarelli-Parisi equations and convoluted with the
hard-scattering cross sections. Notice that, in this approach, the
heavy-quark mass enters the calculation only via the starting
conditions of the PDF's and PFF's.

In this framework, the large logarithmic terms are resummed in the
following way.  The would-be mass singularities
$\propto\ln\left(p_\perp^2/m^2\right)$ are split into two parts.  One
part $\propto\ln\left(p_\perp^2/\mu^2\right)$, where $\mu$ is the
factorization scale, appears in the hard-scattering cross sections,
which have no dependence on $m$.  This part may be eliminated by
choosing $\mu\sim p_\perp$.  The other part
$\propto\ln\left(\mu^2/m^2\right)$ is absorbed into the PFF's.  The
large $\ln\left(\mu^2/\mu_{0}^{2}\right)$, with $\mu_0\sim m$ and
$\mu\sim p_\perp$, is implemented via the evolution equations, and
therefore these large logarithms are resummed.  The residual terms
$\propto\ln\left(m^2/\mu_0^2\right)$ connected with the starting
condition in Eq.~(\ref{eqfrag}) are treated at fixed order in
perturbation theory.

\section{Results}

In the following, we collect our results for the cross section
$\mbox{d}^2\sigma/\mbox{d}y\,\mbox{d}p_\perp^2$ for three cases of
particular interest:
$(i)$ LEP2 with $\sqrt s=175$~GeV,
$(ii)$ Next Linear Collider (NLC) with
$\sqrt s=500$~GeV assuming the TESLA design, and
$(iii)$ NLC with $\sqrt s=500$~GeV operated in the $\gamma\gamma$ mode
implemented by backscattering of laser light on the $e^+$ and $e^-$
beams.

We start with our analysis relevant for LEP2.  The quasi-real-photon
spectrum is described in the Weizs\"acker-Williams approximation (WWA)
by the formula \cite{frix}
\begin{equation}
\label{wwa}
f_\gamma(x)={\alpha\over2\pi}\left\{
{1+(1-x)^2\over x}\ln{E^2\theta_c^2(1-x)^2+m_e^2x^2\over m_e^2x^2}
+2(1-x)\left[{m_e^2x\over E^2\theta_c^2(1-x)^2+m_e^2x^2}-{1\over x}\right]
\right\},
\end{equation}
where $x=E_\gamma/E_e$ is varied over the full range allowed by
kinematics and $\theta_c$ is the maximum angle under which the
outgoing electrons (positrons) are tagged.  In our LEP2 analysis, we
choose $\theta_c=30$~mrad.  All calculations are performed at NLO in
the $\overline{\mbox{MS}}$ renormalization and factorization scheme
using the two-loop formula for $\alpha_s$.  The DR and RR cross
sections are calculated using the photon PDF's of Gl\"uck, Reya, and
Vogt \cite{grv} transformed to the $\overline{\mbox{MS}}$ scheme.  The
renormalization and factorization scales are set to $\mu
=\sqrt{p_\perp^2 + m^2}$.  It is clear that in the massive scheme only
three flavours are active in the initial state and in the evaluation
of $\alpha_s$, whereas in the massless scheme also the charm
distribution in the photon contributes to charm quark production,
and $\alpha_s$ is calculated using four active flavours.  In the
massive calculation and in the PFF's, we set $m=1.5$~GeV.

In Figs.~1a--d, we show the DD, DR, and RR contributions to the cross
section $\mbox{d}^2\sigma/\mbox{d}y\,\mbox{d}p_\perp^2$ as a function
of $p_\perp$ for rapidity $y=0$ and their sum, respectively, both in
the massless and massive schemes.  However, one should bear in mind
that, beyond LO, the separation into the DD, DR, and RR channels
depends on the factorization scheme and scale and has no direct
physical meaning. [As we have discussed above the DD channel in the
massive scheme carries unambigous physical meaning.]  Nevertheless, we
consider these channels separately in order to assess their relative
importance.  As may be seen from these figures, the cross section in
the considered $p_\perp$ range is dominated by the DD contribution.
For the DD and DR components, the massless and massive calculations
are very similar, whereas in the RR case the massive prediction is at
least one order of magnitude smaller than the massless one.  Comparing
the two schemes, we see that the difference is somewhat larger for the
DD contribution than for the DR one.  In both cases, the massive cross
section exceeds the massless one, which is still the case for the
total sum.  The discrepancy increases for increasing $p_\perp$.  The
situation for the DR and RR contributions is comparable to what we
have observed for photoproduction of charm quarks at HERA
\cite{kkks,CG95}.  However, due to the DD channel, which does not
exist in photoproduction at HERA, the total $\gamma\gamma$ results in
the massless and massive schemes exhibit a pattern somewhat different
from \cite{kkks,CG95}. In total, we observe that the massless
prediction for the total sum is smaller than the massive one, in
particular for large $p_\perp$.  We attribute this to the presence of
higher-order leading-logarithmic terms in the PFF's incorporated in
the massless approach, which soften the $p_\perp$ distribution of the
heavy quark.

We now repeat the analysis of Fig.~1 for the $y$ spectrum at
$p_\perp=10$~GeV and display the DD, DR, and RR contributions as well
as the total sum in Fig.~2a--d, respectively.  Comparing the results
in the massive and the massless approaches, we observe a pattern
similar to the $p_\perp$ distribution.  For given $p_\perp$, the
kinematically allowed $y$ range in the massless theory is larger than
that in the massive theory.  This does not show up in Figs.~2a, b, and
d due the influence of the PFF's.  Moreover, at the edges of phase
space, where only soft-gluon emission is allowed, the PFF's are not
reliable due to the missing resummation of Sudakov terms.  In the
region of large cross section, the shapes for the two approaches are
very similar.

Next, we consider the predictions for the NLC with TESLA architecture.
It is well known \cite{orange} that, at the NLC, photons are produced
not only by bremsstrahlung but also via synchrotron radiation emitted
by one of the colliding bunches in the field of the opposing one.
This phenomenon is called beamstrahlung.  The details of the
beamstrahlung spectrum crucially depend on the design and operation of
the NLC.  For certain NLC concepts, beamstrahlung can jeopardize the
overall physics potential due to severe smearing and lowering of the
available centre-of-mass energy. In our study, we select the TESLA
design, where the unwanted effects of beamstrahlung are reduced to an
almost unnoticeable level.  We coherently superimpose the WWA and
beamstrahlung spectra.  We compute the WWA spectrum from
Eq.~(\ref{wwa}) with $\theta_c=175$~mrad and the beamstrahlung
spectrum from the expression given in \cite{orange}, with parameters
$\Upsilon_{\mathrm eff}=0.039$ and $\sigma_z=0.5$~mm \cite{tesla}.
The $p_\perp$ distributions for $y=0$ and $\sqrt s=500$~GeV are shown
in Figs.~3a--d, again for the DD, DR, and RR components, and their
sum, respectively.  Apart from an overall enhancement due to the
increased WWA logarithm, the $p_\perp$ spectra exhibit features very
similar to the LEP2 case, including the relation between the massless
and massive approaches.  The corresponding $y$ spectra are shown in
Figs.~4a--d.  Due to the admixture of beamstrahlung, their shape
differs from that of the pure WWA case in Figs.~2a--d.  This
difference is most pronounced in the DD contribution, shown in
Fig.~4a, which also governs the shape of the total sum.

To achieve the highest possible photon energies with large enough
luminosity, it has been proposed to convert the NLC into a
$\gamma\gamma$ collider via backscattering of the $e^+$ and $e^-$
beams on high-energetic laser light \cite{gkst}.  The corresponding
photon spectrum is given by \cite{gkst}
\begin{displaymath}
f_\gamma(x)=\frac{1}{G(\kappa)}\left(1-x+\frac{1}{1-x}
           -\frac{4x}{\kappa(1-x)}+\frac{4x^2}{\kappa^2(1-x)^2}\right),
\end{displaymath}
with
\begin{equation}
G(\kappa)=\left(1-\frac{4}{\kappa}-\frac{8}{\kappa^2}\right)\ln(1+\kappa)
         +\frac{1}{2}+\frac{8}{\kappa}-\frac{1}{2(1+\kappa)^2}\, .
\end{equation}
Notice that this spectrum extends only up to $x_{\mbox{\scriptsize
    max}}=\kappa/(1+\kappa)$.  In order to avoid the production of
$e^+e^-$ pairs in the collisions of the primary laser photons and the
high-energetic back-scattered photons, one needs to arrange the
experimental set-up so that $\kappa\;\simlt\; 4.83$.  In our analysis,
we choose $\kappa=4.83$, so that $x_{\mbox{\scriptsize max}}=83\%$.
In Figs.~5a--d, we show the $p_\perp$ distributions for $y=0$ due to
the DD, DR, and RR channels, and the total sum, respectively.
Compared to the previous cases, the relative magnitudes of the DD, DR,
and RR components have changed.  At small $p_\perp$, around 5~GeV say,
the massive cross section is dominated by the DR component, which
makes up approximately 60\% of the total sum, while the DD component
is completely negligible.  At the same $p_\perp$ values in the
massless case, RR is the largest component, being 65\% of the full
result.  The DD contribution is again negligible.  The massless cross
section is only slightly larger than the massive one. At large
$p_\perp$, at around 20~GeV, the situation is different.  In the
massless case, the DD, DR, and RR components all have the same order
of magnitude.  In the massive case, the RR contribution is small, and
the total cross section is built up by the DR and DD contributions
approximately in the ratio 3:1.  Looking at Fig.~5d, we see that
$p_\perp$ distributions in the two schemes almost coincide.  However,
the massless result falls off somewhat more strongly with $p_\perp$
increasing.  As may be expected, in both schemes, the DD contributions
are insignificant in the low-$p_\perp$ range but become important for
high $p_\perp$.  The corresponding $y$ spectra for $p_\perp=10$~GeV
are plotted in Figs.~6a--d.  As far as the relative importance of the
individual contributions in the two schemes is concerned, we recognize
the same pattern as in Figs.~5a, b, and c, which refer to $y=0$.  The
line shapes are completely different from those encountered in the
TESLA case.  The $y$ spectrum of the total sum peaks near the
phase-space boundaries.  This is caused by the different photon
spectrum, which is now peaked at the upper edge.  By contrast, the
numerically small DD contribution is almost $y$ independent.  Although
the total sums in the massless and massive calculations have different
decompositions, they nevertheless agree very well with each other over
the full $y$ range.

Finally, we investigate the scale dependence of our results.  To this
end, we introduce a dimensionless scale parameter $\xi$ and set all
scales equal to $\xi\sqrt{m^2+p_\perp^2}$.  In Fig.~7, we plot, versus
$\xi$, the massless and massive cross sections at $y=0$ and
$p_\perp=20$~GeV for LEP2, TESLA, and Laser spectrum.  For
$0.5<\xi<2$, we observe minor scale variations, below $15\%$.  As
expected, the scale dependence is somewhat reduced in the massless
cross section.  We checked that the situation is similar for
$p_\perp=10$~GeV.

\section{Conclusions}

In this paper, we compared two approaches, the massless and massive
schemes, for calculating inclusive charm quark production in three
different arrangements of $\gamma\gamma$ reactions.  The cross
sections were computed at NLO, with PFF's included in the massless
case.  Comparing the massless and massive results in the DD, DR, and
RR channels, we found essential differences, which are compensated to
a large extent in their sums.

In all three arrangements, the cross section at large $p_\perp$ was
found to be somewhat larger in the massive case.  This is in line with
the findings of \cite{CG94} for bottom hadroproduction and of
\cite{CG95} for photoproduction at HERA.  In the latter case, too, it
was observed that, in the massless approximation, the direct and
resolved contributions, which in our case correspond to the DR and RR
contributions, are comparable.  On the other hand, the resolved
contribution of the massive calculation was found to be suppressed
\cite{kkks,CG95}, which also agrees with our $\gamma\gamma$ results.

Concerning the small $p_\perp$ region, we notice that the total
massless result usually overshoots the massive one. This is due to the
missing mass terms in the massless approach, which render it less
reliable in this region.  This effect is enhanced with respect to what
was observed in \cite{CG95}, for the following two reasons: In this
work, we use massless kinematics, whereas, in \cite{CG95}, massive
kinematics was employed. Moreover, in \cite{CG95}, the photon PDF set
ACFGP--ho (mc) \cite{acfgp} was used.  In contrast to GRV used here,
in this set, the massive Bethe-Heitler formula is built in for the
starting condition of the charm PDF of the photon.  This had the
effect that, especially in the low-$p_\perp$ region, the cross section
was reduced.

In conclusion, we have found that, in the region of common validity,
the massive and the massless approaches yield comparable results and
display similar scale dependences. In the large $p_\perp$ region, the
massless approach predicts a lower cross section, due to the
resummation of $\alpha_s\ln(p_\perp^2/m^2)$ terms, and leads to a
reduced sensitivity to the choice of the renormalization and
factorization scales.

\bigskip
\centerline{\bf ACKNOWLEDGMENTS}
\smallskip\noindent
We thank P. Nason for providing us with a computer program for the
evolution of the PFF's. One of us (G.K.) thanks the Theory Group of
the Werner-Heisenberg-Institut for the hospitality extended to him
during a visit when this paper was finalized.

%%%%%%%%%%%%%%%%%%%%%%%%%%%%%%%%%%%%%%%%%%%%%%%%%%%%%%%%%%%%%%%%%%%%%%%%%%%%%%

\newpage

\centerline{\bf FIGURE CAPTIONS}

\noindent
{\bf Fig.~1a--d}
Inclusive cross section
$\mbox{d}^2\sigma/\mbox{d}y\,\mbox{d}p_\perp^2$ for
$e^+e^-\to e^+e^-c/\bar c+X$ as a function of $p_\perp$ for
$\protect\sqrt s=175$~GeV
(LEP2) and $y=0$ in the massless (solid lines) and massive (dashed lines)
schemes: $(a)$ DD, $(b)$ DR, $(c)$ RR, and $(d)$ total sum.

\noindent
{\bf Fig.~2a--d}
Inclusive cross section
$\mbox{d}^2\sigma/\mbox{d}y\,\mbox{d}p_\perp^2$ for
$e^+e^-\to e^+e^-c/\bar c+X$ as a function of $y$ for $\sqrt s=175$~GeV (LEP2)
and $p_\perp=10$~GeV in the massless (solid lines) and massive (dashed lines)
schemes: $(a)$ DD, $(b)$ DR, $(c)$ RR, and $(d)$ total sum.

\noindent
{\bf Fig.~3a--d}
Same as Fig.~1a--d for
$\sqrt s=500$~GeV (NLC with WWA plus beamstrahlung).

\noindent
{\bf Fig.~4a--d}
Same as Fig.~2a--d for
$\sqrt s=500$~GeV (NLC with WWA plus beamstrahlung).

\noindent
{\bf Fig.~5a--d}
Same as Fig.~1a--d for
$\sqrt s=500$~GeV (NLC with laser spectrum).

\noindent
{\bf Fig.~6a--d}
Same as Fig.~2a--d for
$\sqrt s=500$~GeV (NLC with laser spectrum).

\noindent
{\bf Fig.~7a--c}
Scale dependence of
$\mbox{d}^2\sigma/\mbox{d}y\,\mbox{d}p_\perp^2$ for
$e^+e^-\to e^+e^-c/\bar c+X$ at $y=0$ and $p_\perp=20$~GeV
in the massless (solid lines) and massive (dashed lines) schemes:
$(a)$ $\sqrt s=175$~GeV (LEP2),
$(b)$ $\sqrt s=500$~GeV (NLC with WWA plus beamstrahlung), and
$(c)$ $\sqrt s=500$~GeV (NLC with laser spectrum).
The renormalization and factorization scales are identified and set equal to
$\xi\sqrt{p_\perp^2 + m^2}$.

%\end{document}

\newpage

\vspace*{4.0cm}
\hspace*{-2.5cm}
\begin{turn}{-90}%
\epsfxsize=13cm \epsfbox{fig.1}
\end{turn}
\vspace*{1.5cm}

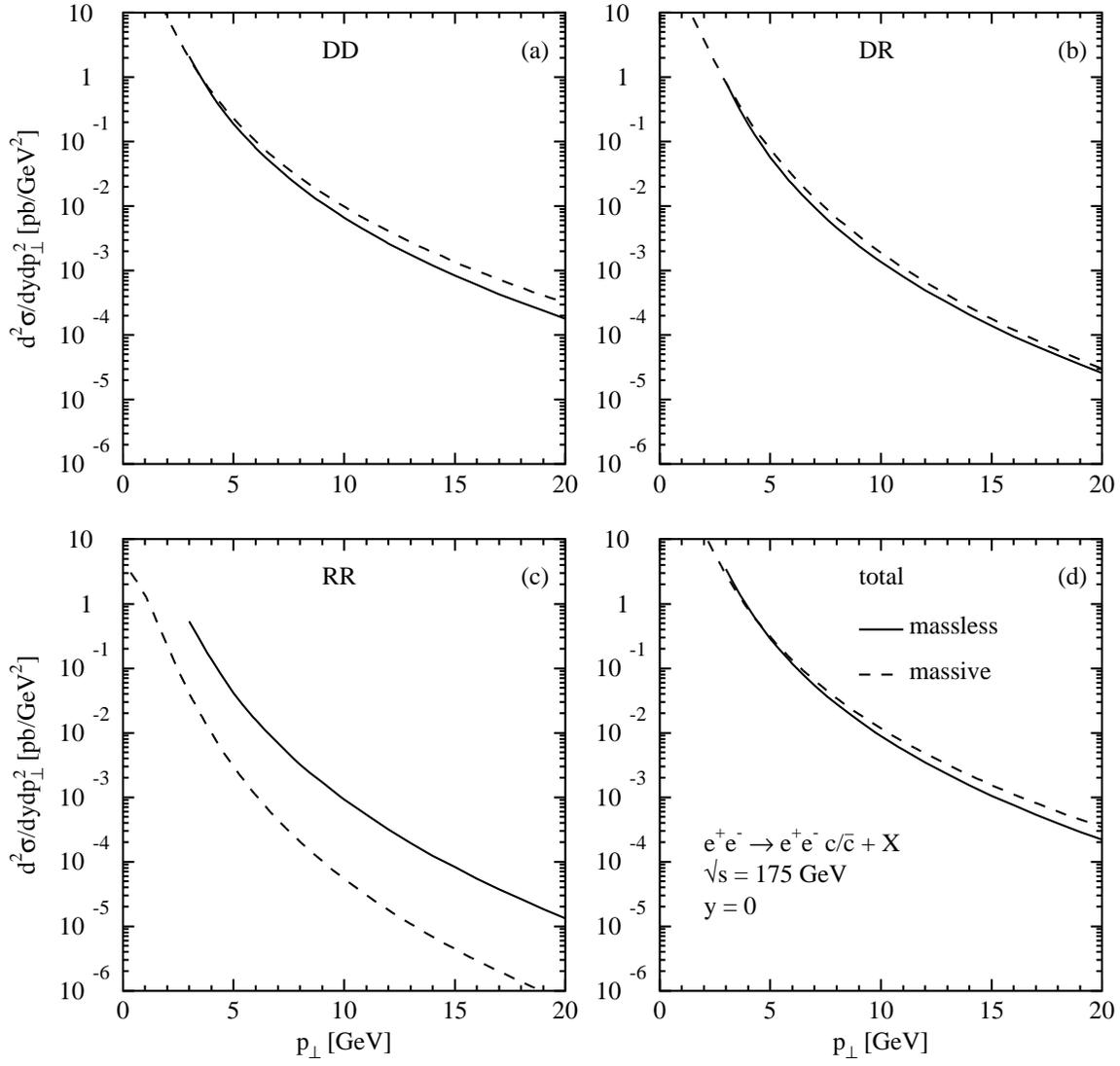
\centerline{\bf Fig.~1}

\newpage

\vspace*{4.0cm}
\hspace*{-2.5cm}
\begin{turn}{-90}%
\epsfxsize=13cm \epsfbox{fig.2}
\end{turn}
\vspace*{1.5cm}

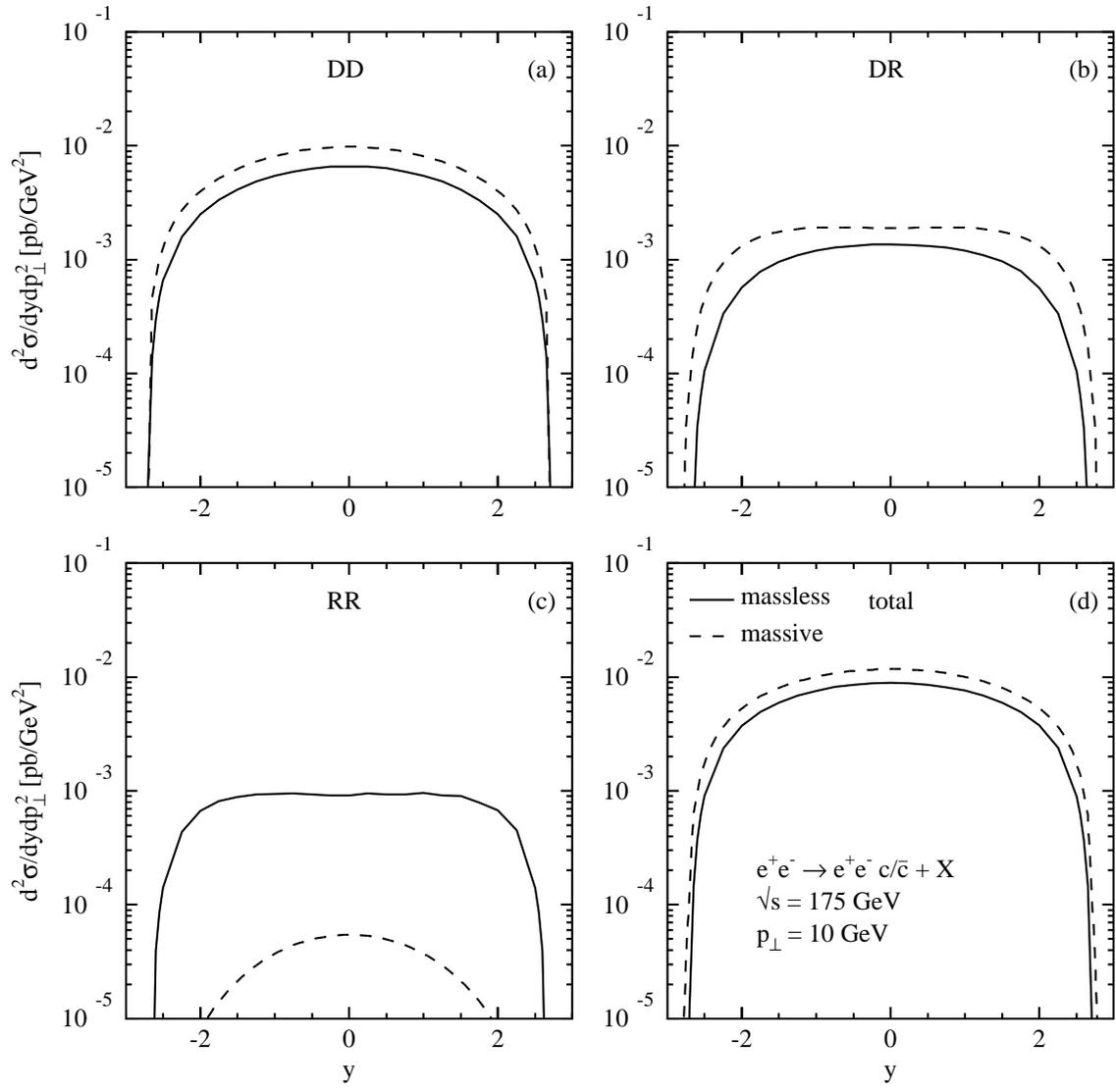
\centerline{\bf Fig.~2}

\newpage

\vspace*{4.0cm}
\hspace*{-2.5cm}
\begin{turn}{-90}%
\epsfxsize=13cm \epsfbox{fig.3}
\end{turn}
\vspace*{1.5cm}

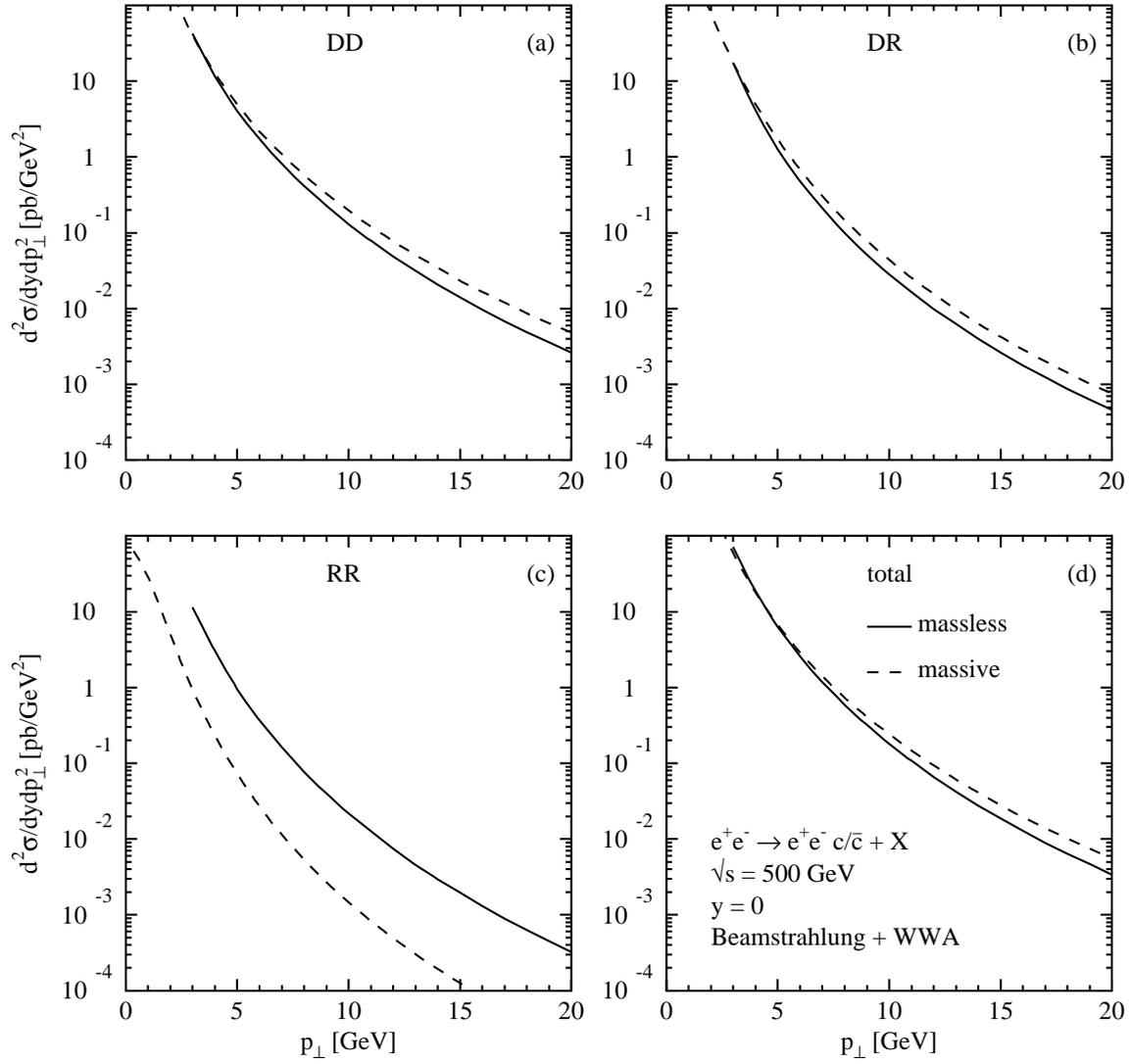
\centerline{\bf Fig.~3}

\newpage

\vspace*{4.0cm}
\hspace*{-2.5cm}
\begin{turn}{-90}%
\epsfxsize=13cm \epsfbox{fig.4}
\end{turn}
\vspace*{1.5cm}

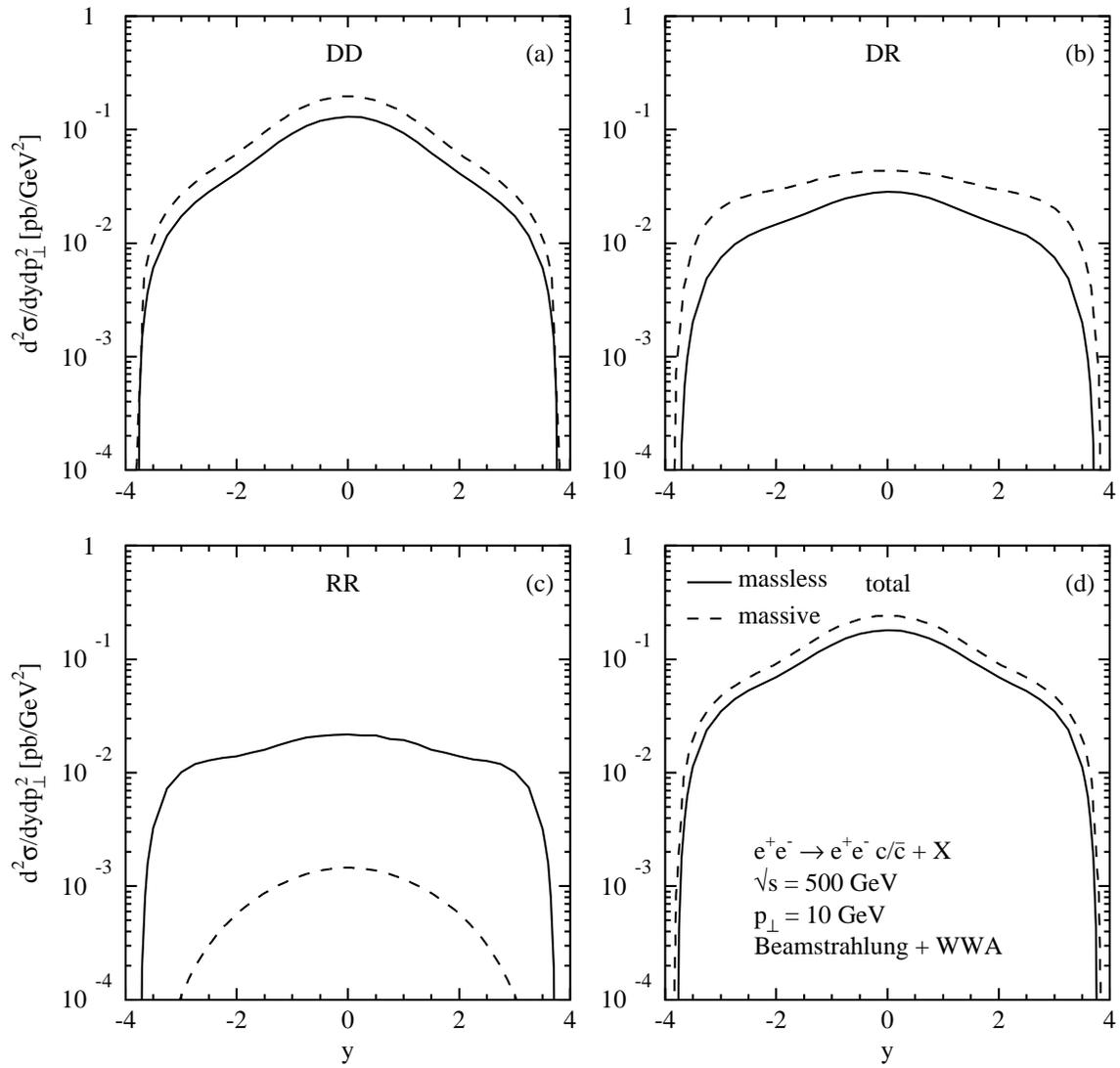
\centerline{\bf Fig.~4}

\newpage

\vspace*{4.0cm}
\hspace*{-2.5cm}
\begin{turn}{-90}%
\epsfxsize=13cm \epsfbox{fig.5}
\end{turn}
\vspace*{1.5cm}

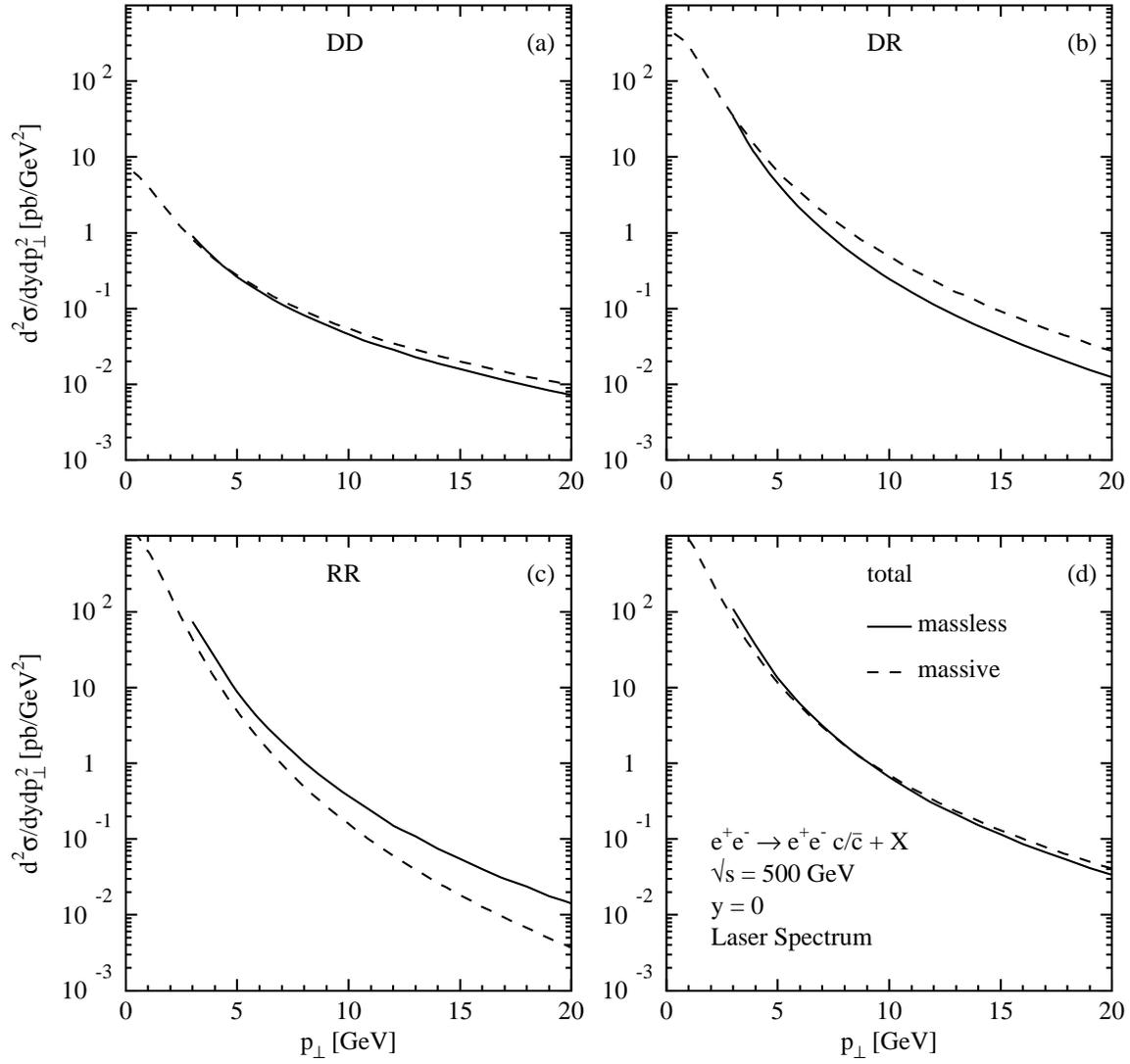
\centerline{\bf Fig.~5}

\newpage

\vspace*{4.0cm}
\hspace*{-2.5cm}
\begin{turn}{-90}%
\epsfxsize=13cm \epsfbox{fig.6}
\end{turn}
\vspace*{1.5cm}

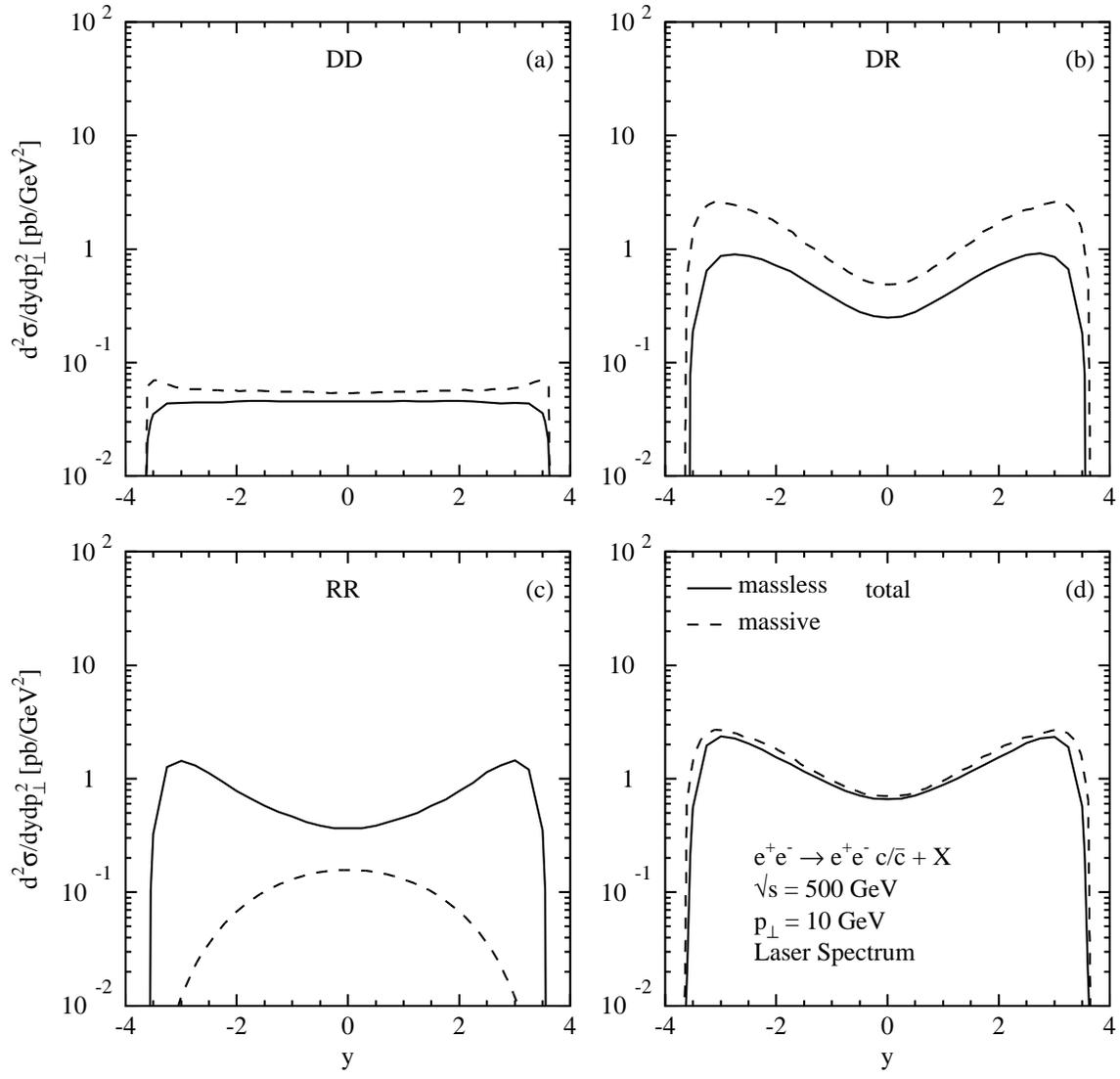
\centerline{\bf Fig.~6}

\newpage

\vspace*{4.0cm}
\hspace*{-2.5cm}
\begin{turn}{-90}%
\epsfxsize=13cm \epsfbox{fig.7}
\end{turn}
\vspace*{1.5cm}

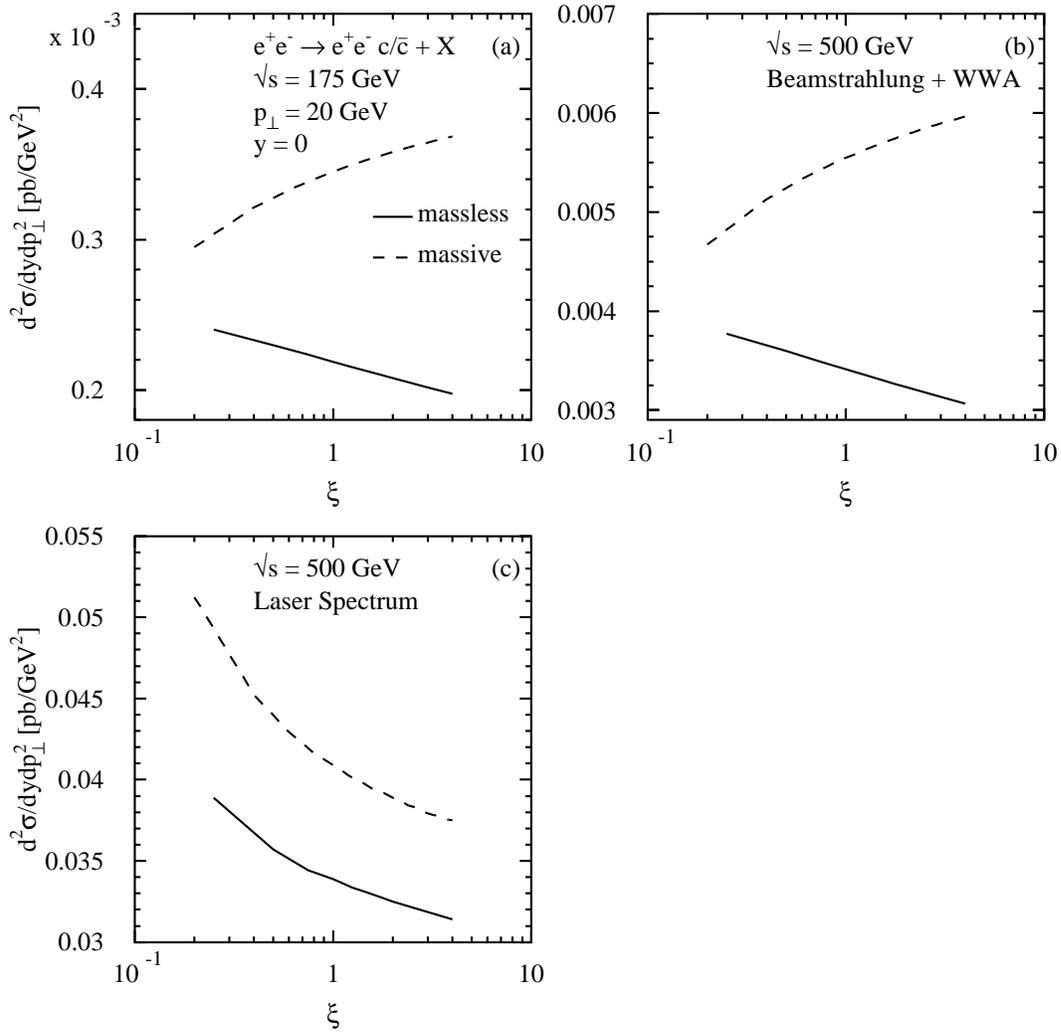
\centerline{\bf Fig.~7}

\end{document}